\documentclass[twocolumn,showpacs,superscriptaddress,preprintnumbers,prd]{revtex4}
\usepackage{graphicx}
\usepackage{amsmath}

\begin{document}
\preprint{SAGA-HE-200-03}
\preprint{KUNS1885}

\title{\Large \bf Determination of polarized parton distribution functions \\
                  and their uncertainties}
\author{M. Hirai}
\email[E-mail: ]{mhirai@rarfaxp.riken.jp}
\affiliation{Radiation Laboratory, 
RIKEN (The Institute of Physical and Chemical Research) \\
Wako, Saitama 351-0198, JAPAN}
\author{S. Kumano}
\email[E-mail: ]{kumanos@cc.saga-u.ac.jp}
\affiliation{Department of Physics, Saga University,
         Saga, 840-8502, Japan}
\author{N. Saito}
\email[E-mail: ]{saito@nh.scphys.kyoto-u.ac.jp}
\affiliation{Department of Physics, Kyoto University, 
         Kyoto, 606-8502, Japan}
\collaboration{Asymmetry Analysis Collaboration}

\date{\today}
\begin{abstract}
We investigate the polarized parton distribution functions
(PDFs) and their uncertainties by using the world data on the spin
asymmetry $A_1$. The uncertainties of the polarized PDFs are estimated
by the Hessian method. The up and down valence-quark distributions are
determined well. However, the antiquark distributions have large uncertainties
at this stage, and it is particularly difficult to fix the gluon distribution.
The $\chi^2$ analysis produces a positively polarized gluon distribution,
but even $\Delta g(x)=0$ could be allowed according to our uncertainty
estimation. In comparison with the previous AAC (Asymmetry Analysis
Collaboration) parameterization in 2000, accurate SLAC-E155 proton data are
added to the analysis. We find that the E155 data improve the determination of
the polarized PDFs, especially the polarized antiquark distributions. 
In addition, the gluon-distribution uncertainties are reduced due to
the correlation with the antiquark distributions. We also show the global
analysis results with the condition $\Delta g(x)=0$ at the initial scale,
$Q^2=1$ GeV$^2$, for clarifying the error correlation effects with
the gluon distribution.
\end{abstract}

\maketitle

\section{Introduction}

The spin structure of the nucleon has been investigated extensively since
the discovery of the EMC spin effect \cite{EMC}. Despite a naive expectation
that the nucleon spin is carried by quarks, the experimental result indicated
that only a small fraction is carried by the quarks. In order to determine
this quark spin content and internal spin structure, the determination of
the first moments of the polarized parton distribution functions (PDFs) is 
necessary. Furthermore, the $x$ and $Q^2$ dependence of these functions
is crucial in the calculation of the polarized cross sections. 
The functions should be determined from global analyses of 
polarized experimental data. Such analyses have been made by several groups 
\cite{DS,GRSV,GS,BT,ABFR,BBPSS,GGR,SMCf,LSS,AAC,GGI,BB}.
Now, there are available data for the spin asymmetry $A_1$ or
the structure function $g_1$ by
polarized deep inelastic scattering (DIS) experiments
\cite{EMC,SMC,HERMES,SLAC,E155p}. These data are valuable especially
for fixing the the polarized valence quark distributions 
$\Delta u_v(x)$ and $\Delta d_v(x)$. The polarized antiquark distributions
are still not well determined from the data. In particular, their flavor
dependence is not shown reliably at this stage. Furthermore,
the polarized gluon distribution cannot be fixed from the polarized DIS
data although there are some constraints.

The above statements describe the current status of 
global analyses qualitatively well; however, we have been longing for 
more quantitative discussion on the uncertainties in the PDFs.  
Recently, uncertainty estimation methods have been developed
for the unpolarized PDFs. Mathematical formulations
of PDF uncertainties are proposed in Refs. \cite{Fmathu,GKK}.
Practical methods are also developed
\cite{CTEQ-l,CTEQ-h,Alehkin,Botje}
and they are actually used in recent unpolarized analyses
\cite{CTEQ6,ZEUS,MRST02}.
Thanks to a large number of experimental data points with 
excellent precision and wide kinematical coverage, 
the unpolarized PDFs are well determined from small $x$ to large $x$
with a reasonable precision so that hadronic cross sections 
can be calculated to a few percent accuracy \cite{CTEQ6,MRST02,STUMP03}.
In the same way, the uncertainties of the polarized PDFs have been
investigated \cite{LSS,BB}. The polarized gluon distribution has large
uncertainties due to a lack of data which are sensitive to the distribution.
However, because polarized experimental projects
are going on, we expect to have better determination of $\Delta g$
in the near future. The determination of the polarized PDF uncertainties
enables the uncertainty estimation of physical observables such as 
scattering cross sections and spin asymmetries.
The uncertainty estimation of the polarized PDFs is valuable
for understanding the present situation and
for clarifying the importance of future experiments.

There are three major purposes in this paper.
First, the uncertainties of the polarized PDFs are investigated for
the AAC (Asymmetry Analysis Collaboration) parameterization \cite{AAC}.
Although there are uncertainty estimations for the polarized PDFs
\cite{LSS,BB}, the results could depend on the parameterized functional form
and the details of the uncertainty estimation method. 
Therefore, it is
important to estimate the PDF uncertainties independently. In particular,
we discuss the large uncertainties of the polarized gluon distribution. 
Second, we investigate the impact of precise SLAC-E155 proton data,
which are not included in the previous AAC analysis,
on the PDF uncertainties, especially on those of the polarized antiquark
and gluon distributions. Third, error correlation is investigated
by a global analysis with $\Delta g(x)=0$ at the initial $Q^2$ point.
We compare its PDF uncertainties with those of the $\Delta g(x) \ne 0$
analysis in order to show error correlation effects.

This paper is organized as follows.
In Sec. \ref{AAC analysis}, we describe the framework of our parameterization
for the polarized PDFs. The Hessian method is explained in Sec. \ref{Hessian}
as an uncertainty estimation method for the PDFs. 
In Sec. \ref{results}, $\chi^2$ analysis results are shown with
the polarized PDF uncertainties.
First, they are compared with the experimental data for the spin
asymmetry $A_1$. Second, obtained polarized PDFs are shown in comparison
with the distributions of the previous AAC version and other parameterization
studies. Third, effects of the SLAC-E155 data are explained, and
the correlation between the antiquark and gluon distribution
uncertainties is discussed. The results are summarized in Sec. \ref{Summary}.

\section{\label{AAC analysis} 
          Parameterization of polarized parton distribution functions}

The major source of information on the polarized PDFs has been 
polarized electron and muon DIS experiments. 
The polarized PDFs are determined by comparing theoretical functions
with the asymmetry $A_1$ data of the polarized DIS experiments 
\cite{E155p,EMC,SMC,HERMES,SLAC}. 
The variable $Q^2$ is given by $Q^2 = -q^2$ with
the momentum transfer $q$, and the scaling variable $x$ is defined
by $x=Q^2/(2p\cdot q)$ with the nucleon momentum $p$.
We selected the data with $Q^2>1$ GeV$^2$
so that perturbative QCD could be applied relatively safely. 
Then, the total number of available data is 399, and they cover
the kinematical region, $0.004<x<0.75$ and $1<Q^2<72$ GeV$^2$.

The spin asymmetry $A_1$ is expressed in terms of
the polarized structure function $g_1$, 
the unpolarized structure function $F_2$, and
the longitudinal-transverse structure function ratio $R$:
\begin{equation}
        A_1(x, Q^2)=\frac{g_1(x, Q^2)}{F_2(x, Q^2)}\,
               2 \, x \, [1+R(x, Q^2)] \, .
\end{equation}
The SLAC analysis results \cite{SLAC-R} are used for the function $R(x,Q^2)$.
The structure function $F_2(x,Q^2)$ is expressed by unpolarized PDFs:
\begin{align}
F_2(x,Q^2) = \sum\limits_{i=1}^{n_f} e_{i}^2 \, x \, &
\bigg\{ C_q(x,\alpha_s) \otimes [ q_i(x,Q^2) +\bar{q}_{i}(x,Q^2) ]
\nonumber \\
     & + C_g(x,\alpha_s) \otimes g (x,Q^2)  \bigg\}.
\label{eqn:f2}
\end{align}
Here, $q(x,Q^2)$, $\bar{q}(x,Q^2)$, and $g(x,Q^2)$ are
the quark, antiquark, and gluon distributions, respectively,
and $C_{q}(x,\alpha_s)$ and $C_{g}(x,\alpha_s)$ are coefficient functions.
The symbol $\otimes$ denotes the convolution integral:
\begin{equation}
f (x) \otimes g (x) = \int^{1}_{x} \frac{dy}{y}
            f\left(\frac{x}{y} \right) g(y) .
\end{equation}
In the same way, the structure function $g_1(x, Q^2)$ is expressed as 
\begin{align}
g_1 (x, & Q^2) = \frac{1}{2}\sum\limits_{i=1}^{n_f} e_{i}^2
   \bigg\{ \Delta C_q(x,\alpha_s) \otimes [ \Delta q_{i} (x,Q^2)
\nonumber \\
   & + \Delta \bar{q}_{i} (x,Q^2) ]
    + \Delta C_g(x,\alpha_s) \otimes \Delta g (x,Q^2) \bigg\},
\end{align}
where $\Delta q(x,Q^2)$, $\Delta \bar{q}(x,Q^2)$, and $\Delta g(x,Q^2)$ 
are the polarized quark, antiquark, and gluon distributions, respectively.
The function $\Delta q$ is defined by
$\Delta q = q^{\uparrow}-q^{\downarrow}$, which indicates
the difference between the distribution of quark with
helicity parallel to that of parent nucleon and the one
with helicity anti-parallel.
The functions $\Delta C_{q}(x,\alpha_s)$ and $\Delta C_{g}(x,\alpha_s)$
are the polarized coefficient functions.

As the polarized PDF at the initial scale $Q_0^2$,
we choose the functional form:
\begin{equation}
        \Delta f(x) = [\delta x^{\nu}-\kappa (x^{\nu}-x^{\mu})] f(x) \, ,
\label{eqn:df}
\end{equation}
where $f(x)$ is the corresponding unpolarized PDF. 
This form is taken for imposing the positivity condition and
for reducing correlations among the parameters.
Optimized PDFs are four distributions: 
$\Delta u_v(x)$, $\Delta d_v(x)$, $\Delta \bar{q}(x)$, and $\Delta g(x)$,
which are defined at $Q_0^2$ by Eq. (\ref{eqn:df}).
The $\delta$, $\nu$, $\kappa$, and $\mu$ are free parameters, which 
are determined by a $\chi^2$ analysis.

In principle, the separation of these quark distributions 
can be arbitrarily chosen. 
For example, alternative choice would be 
$\Delta u^+(x)$, $\Delta d^+(x)$ and $\Delta s^+(x)$, where
\begin{equation} 
  \Delta u^+(x) = \Delta u(x) + \Delta \bar{u}(x) \, , 
\end{equation} 
and similar expressions for $\Delta d^+(x)$ and $\Delta s^+(x)$. 
Here, these quark distributions can be related to ours as
\begin{eqnarray}
  \Delta u^+(x) &=& \Delta u_v(x) + 2 \Delta \bar{q}(x) \, , \nonumber \\
  \Delta d^+(x) &=& \Delta d_v(x) + 2 \Delta \bar{q}(x) \, ,           \\
  \Delta s^+(x) &=& 2 \Delta \bar{q}(x)                 \, . \nonumber
\end{eqnarray}
Here $\Delta \bar{q}(x)$ can simply be interpreted 
as an average of polarized sea-quark distributions. 

Practical difference appears, however, when we apply constraints on the 
first moments of the quark distributions from the axial coupling constants
of octet baryons. By denoting the first moments by 
$\int_0^1 \Delta f(x) dx = \Delta F$, these moments should be 
connected to 
\begin{eqnarray}
\Delta U^+ - \Delta D^+ & = & 1.267  \pm 0.011        \, , \nonumber \\
\Delta U^+ + \Delta D^+ - 2\Delta S^+ & = & 0.585 \pm 0.025  \, .
\end{eqnarray}
These relations can be rewritten by using our separation of quark 
distributions, 
\begin{eqnarray}
\Delta U_v - \Delta D_v & = & 1.267 \pm 0.011  -2 \Delta_2   \, , 
\nonumber \\
\Delta U_v + \Delta D_v & = & 0.585 \pm 0.025  -4 \Delta_3   \, .
\label{E:BetaDecay}
\end{eqnarray}
where
\begin{eqnarray}
\Delta_2 & \equiv &  \Delta \overline{U} - \Delta \overline{D} \, , \nonumber \\ 
\Delta_3 & \equiv &                                 
\frac{\Delta \overline{U} + \Delta \overline{D}}{2} - \Delta \overline{S} \, . 
\end{eqnarray}
Since there is no experimental guidance on the size of 
the first moments of the flavor asymmetric distributions
($\Delta_2$ and $\Delta_3$), 
we continue to neglect them in this paper as was done in our previous one, 
although we are also preparing 
a new calculation with those breaking parameters. 
This point will be discussed later in this section. 

The polarized distributions are numerically evolved to the $Q^2$ points of
experimental data by the DGLAP (Dokshitzer-Gribov-Lipatov-Altarelli-Parisi)
evolution equations \cite{Q2EVOL} in order to calculate $\chi^2$. 
The total $\chi^2$ is defined by
\begin{equation}
\chi^2=\sum_i \frac{[ A_{1, \, i}^{\rm data}(x,Q^2)
                     -A_{1, \, i}^{\rm calc}(x,Q^2) ]^2}
                {[\Delta A_{1, \, i}^{\rm data}(x,Q^2) ]^2} ,
\end{equation}
where $\Delta A_1^{\rm data}$ is the experimental error
including both systematic and statistical errors:
$(\Delta A_1^{\rm data})^2 = (\Delta A_1^{\rm stat})^2
                              +(\Delta A_1^{\rm syst})^2$.
The total $\chi^2$ is minimized
by the CERN subroutine {\tt MINUIT} \cite{minuit}. 

Here, the systematic errors $\Delta A_1^{\rm syst}$ are fully included.
It would be ideal to include uncorrelated and correlated systematic errors 
separately so that we can perform fully consistent uncertainty analysis.
The issue of the correlated errors in the global analysis is indeed
investigated in the recent unpolarized PDF parametrizations
\cite{CTEQ6,ZEUS}. In the polarized PDF analysis, however,
these errors are not listed separately in papers 
and it is very difficult to access such information. 
Because of this incompleteness, 
our analyses overestimate the uncertainty in the PDFs. 

In order to obtain a rough picture of the effects of 
systematic uncertainties, we also performed $\chi^2$ 
calculation with the statistical uncertainties only, which 
resulted in $\sim$20\% increase in the $\chi^2$. 
In average, this increase in $\chi^2$ corresponds to the fact that
quadratic sum of statistical error and full systematic error
is larger by $\sim$10\%. 
By looking at the individual data points, 
the increase in the errors ranges  
from a few percent to 50\%, which is not inconsistent with 
the average picture. From this exercise, we can conclude 
that our uncertainties of the PDFs are overestimated, 
but only by $\sim$10\% in average, 
although we now emphasize the needs of separated systematic errors 
so that we can perform fully consistent uncertainty analyses in future. 

In comparison with the previous AAC analysis \cite{AAC}, the SLAC-E155
proton data are added to the $\chi^2$ analysis.
In order to demonstrate the impact of this new data set,
we used the same configuration with the previous analysis.
The renormalization scheme is the $\overline{\rm MS}$ scheme
in the next-to-leading order (NLO).
The initial scale is set at $Q^2_0= 1\ \rm{GeV}^2$. 
The number of flavor is fixed at $N_f=3$ and
heavy flavor contributions are neglected.

We use the GRV98 NLO parameter set as the unpolarized PDFs \cite{GRV98}.
Even if other unpolarized PDFs, for example CTEQ6 \cite{CTEQ6}, are used,
the results do not change significantly. In particular, the polarized
gluon distribution is modified; however, it is well within the uncertainties
for $\Delta g (x)$ in Sec. \ref{PDF}. We should also mention that
unpolarized PDF uncertainties are not included in our analysis
for estimating the polarized PDF uncertainties.
The value of $\Lambda_{QCD}$ is taken from the GRV unpolarized PDF
analysis: $\Lambda^{N_f=3}_{QCD}=299$ MeV.
Unpolarized experimental data are generally more accurate than polarized
ones, so that $\Lambda_{QCD}$ had better be determined by the unpolarized
analysis.

In the following, we discuss two important constraints,
the positivity and flavor-symmetric conditions, on the polarized PDF
determination.

\subsection{The positivity condition}

The positivity condition means that the magnitude of a polarized
cross section should be smaller than the corresponding unpolarized one:
$|\Delta \sigma| \le \sigma$. In the leading order (LO), this relation
indicates $|\Delta f(x)| \le f(x)$ because probabilistic interpretation
can be applied for the parton distributions.
However, the condition is not strictly satisfied in the NLO because of
higher-order corrections \cite{Positivity}. The correction due to
the coefficient functions is small in the limit $x \to 1 $.
Therefore, the positivity condition $|\Delta f(x)| \le f(x)$ could
be used practically for constraining the polarized PDFs at large $x$.

If this condition is not taken into account in the $\chi^2$ analysis,
it tends to be violated significantly in the polarized antiquark and gluon
distributions: $[|\Delta f(x)|/f(x)]_{x\to1} \gg 1$.
It could lead to an unphysical cross section: $|\Delta \sigma| > \sigma$.
This unfavorable behavior comes from the lack of accurate experimental
data in the large-$x$ region. Furthermore, experimental data indicate
that the spin asymmetry $A_1$ increases monotonically as a function
of $x$ in the large-$x$ region. It easily leads to unphysical results
without the positivity constraint.
Therefore, the positivity condition for the polarized PDFs
is practically a useful constraint for avoiding the unphysical results,
and we decided to impose this condition in the $\chi^2$ analysis.
The condition restricts the range of the parameter $\delta$ 
which controls the large-$x$ behavior of the polarized PDFs: 
$-1 \le \delta \le 1 $.

\subsection{Flavor symmetric antiquark distributions}

It is now known that unpolarized antiquark distributions,
$\bar u$, $\bar d$, and $\bar s$, are different from lepton scattering
and Drell-Yan experiments \cite{flavor3}. There are model explanations,
for example, by meson clouds, chiral soliton, and Pauli exclusion.
These models are extended to the polarized antiquark distributions.
There are available experimental data which may indicate
the polarized flavor dependence \cite{DS,f-fit,f-hermes}; however, they are
not accurate enough to find the difference between $\Delta \bar u$ and
$\Delta \bar d$. Therefore, there is no reliable data for fixing the
difference between the polarized antiquark distributions, and
the determination of the flavor asymmetric distributions still depend on 
separation models \cite{GRSV,LSS,DS}.

Even in the alternative quark separation, $\Delta u^+(x)$, 
$\Delta d^+(x)$ and $\Delta s^+(x)$, we cannot address these
specific questions. 
In future, the flavor dependence of the polarized antiquark
distributions will be investigated, for example, at RHIC 
by $W$ production \cite{rhic}. If these data become
available, it makes sense to introduce the flavor dependent
parameters into the $\chi^2$ analysis.

With the flavor symmetric assumption, {\it i.e.} $\Delta_2 = \Delta_3 =0$,
the first moments of $\Delta u_v$ and $\Delta d_v$ are fixed from 
Eq. (\ref{E:BetaDecay}): $\Delta U_v=0.926$ and $\Delta D_v=-0.341$.

\section{\label{Hessian} PDF uncertainty}

The uncertainties of the parton distributions are estimated by the Hessian
method, which has been used as a general statistical method for
estimating errors. The uncertainties come from measurement errors
in the global PDF analysis. 

The parameters, {\it e.g.} $\delta$, $\nu$, $\kappa$, and $\mu$ for
each distribution in our analysis, are denoted $a_i$
($i$=1, 2, $\cdot \cdot \cdot$, $N$), where $N$ is the number of
optimized parameters.  Expanding $\chi^2$ around the minimum point $\hat a$
and keeping the leading quadratic terms, we have
\begin{equation}
        \Delta \chi^2 \equiv \chi^2(\hat{a}+\delta a)-\chi^2(\hat{a})
        =\sum_{i,j} H_{ij}\delta a_i \delta a_j \ ,
        \label{eq:ellipse}
\end{equation}
where the Hessian $H_{ij}$ is the second derivative matrix 
in the function $\chi^2(a)$. The first derivative terms do not exist 
because they vanish at the minimum point.

For estimating the PDF uncertainty, we should evaluate a confidence region
of the multivariate normal distribution for optimized-parameter errors.
Equation (\ref{eq:ellipse}) indicates the local behavior around $\hat{a}$,
and the confidence region could be identified by an ellipsoid which is
defined by $\Delta \chi^2$.
Assuming parabolic curves for the function $\chi^2(a)$, we can vary
the ellipsoid to an arbitrary confidence level by choosing $\Delta \chi^2$.
In our estimation, the $\Delta \chi^2$ value is obtained 
by the following procedure. 

The confidence level $P$ for the normal distribution
with $N$ degrees of freedom is identified with 
the one for the $\chi^2$ distribution:
\begin{equation}
        P=\int_0^{\Delta \chi^2} \frac{1}{2\ \Gamma(\frac{N}{2})} 
        \left(\frac{S}{2}\right)^{\frac{N}{2}-1} 
               \exp\left(-\frac{S}{2} \right) dS \ ,
        \label{eq:dchi2}
\end{equation}
where $\Gamma(m)$ is the Gamma function. 
It can be chosen to be the probability of one-$\sigma$-error range
of the normal distribution ($P=0.6826$)  for our study in order to
compare with the experimental errors. 
In the case of one parameter ($N=1$), we obviously
have $\Delta \chi^2=1$ from Eq. (\ref{eq:dchi2}). Therefore,
$\Delta \chi^2=1$ could be simply used for calculating the uncertainty
if the parameter number is one. However, the parton distributions
are provided with many parameters, so that the $\Delta \chi^2$ value
should be re-evaluated. For example, the parameter number is eleven ($N=11$)
in our new polarized PDF analysis, and it leads to $\Delta \chi^2=12.647$.

The uncertainty of a parton distribution $F(x,\hat{a})$ with respect to
the optimized parameters $\hat{a}$ is then calculated by using the Hessian
matrices and assuming linear error propagation:
\begin{equation}
        [\delta F(x)]^2=\Delta \chi^2 \sum_{i,j}
          \left( \frac{\partial F(x,\hat{a})}{\partial a_i}  \right)
          H_{ij}^{-1}
          \left( \frac{\partial F(x,\hat{a})}{\partial a_j}  \right) \ .
        \label{eq:erroe-M}
\end{equation}
For the PDF uncertainty estimation, we can analytically calculate
the gradient terms $\partial F(x,\hat{a})/\partial a_i$ at the initial
scale $Q^2_0$. For the estimation at arbitrary $Q^2$, 
each gradient term is evolved by the DGLAP evolution kernel,
and then the PDF uncertainties $\delta \Delta f(x,Q^2)$ are calculated.
The uncertainties of the polarized structure functions
$g_1^p$, $g_1^n$, and $g_1^d$ are calculated by the convolution
integrals of the PDF gradient terms with the coefficient functions.

\section{\label{results} Results and discussion}

\begin{table}[b]
\caption{\label{T:chi2}
Numbers of the $A_1$ data with $Q^2>1$ GeV$^2$ and
$\chi^2$ values are listed.
The notations $p$, $n$, and $d$ indicate proton, neutron, and deuteron,
respectively.
}
\begin{ruledtabular}
\begin{tabular}{ccc} 
data set       & No. of data & $\chi^2$ \\
\hline
EMC ($p$)      & 10  &  4.5  \\
SMC ($p$)      & 59  & 54.0  \\
E130 ($p$)     & 8   &  4.9  \\
E143 ($p$)     & 81  & 61.1  \\
E155 ($p$)     & 24  & 24.2  \\
HERMES ($p$) \ & 19  & 19.1  \\
\hline                
SMC ($d$)      & 65  & 56.5  \\
E143 ($d$)     & 81  & 93.6  \\
E155 ($d$)     & 24  & 20.3  \\
\hline                
E142 ($n$)     & 8   &  2.6  \\
E154 ($n$)     & 11  &  3.6  \\
HERMES ($n$)   & 9   &  2.2  \\
\hline
total          & 399 & 346.5 
\end{tabular}
\end{ruledtabular}
\end{table}

\begin{table*}
\caption{\label{T:NLO}
Obtained parameters by the NLO analysis (AAC03).
}
\begin{ruledtabular}
\begin{tabular}{ccccc} 
distribution \  & $\delta$             & $\nu$
                & $\kappa$             & $\mu$  \\
\hline
$\Delta u_v$    &    0.975 $\pm$ 0.099 &  0.000 (fixed)     
                &    0.601             &  1.095 $\pm$ 0.266 \\
$\Delta d_v$    & $-$1.000 $\pm$ 0.377 &  0.000 (fixed)     
                & $-$0.721             &  1.318 $\pm$ 0.466 \\
$\Delta \bar q$ &    1.000 $\pm$ 0.994 &  1.014 $\pm$ 0.182
                &$-$ 90.96 $\pm$ 13.57 &  1.000 (fixed)     \\
$\Delta g$      & $-$1.000 $\pm$ 3.959 &  2.248 $\pm$ 1.089 
                &    254.2 $\pm$ 180.7 &  2.217 $\pm$ 2.172
\end{tabular}
\end{ruledtabular}
\end{table*}

We report our analysis results. Because one of the major purposes is
to show the PDF uncertainties, we analyzed the data only for the NLO set.
In our new analysis, the best fit is obtained with
$\chi^2\ (/d.o.f.)=346.5\ (0.893)$. The $\chi^2$ contributions from
all the used data sets are listed in Table \ref{T:chi2}, and optimized
parameters are summarized in Table \ref{T:NLO}.

In the previous AAC version (AAC00) \cite{AAC}, we found that
an antiquark parameter $\mu_{\bar q}$, which determines the functional
behavior of $\Delta \bar q$ at small $x$, cannot be fixed from the data
because of the lack of small-$x$ data. Therefore, we fixed the parameter at
$\mu_{\bar q}=1$ in our new analysis. 
Other four parameters are also fixed. The parameter $\nu_{u_v}$
and $\nu_{d_v}$ tended to stop at the positivity limit, so that these
parameters are finally fixed. The parameters $\kappa_{u_v}$
and $\kappa_{d_v}$ are determined by the first moments
$\Delta u_v$ and $\Delta d_v$ from semileptonic data
with the assumption for the flavor symmetric antiquark distributions.
The difference from the AAC00 NLO-2 analysis is the addition of
the SLAC-E155 data. In order to discuss the influence of such accurate
experimental data on the polarized PDF analysis, new analysis results are
compared to the AAC00-NLO-2 in following subsections. 
The total number of the optimized parameters is eleven, so that
the uncertainty is estimated by $\Delta \chi^2=12.647$
as explained in Sec. \ref{Hessian}.

\subsection{\label{spinA} Spin asymmetries}

\begin{figure}
      \includegraphics*[width=60mm]{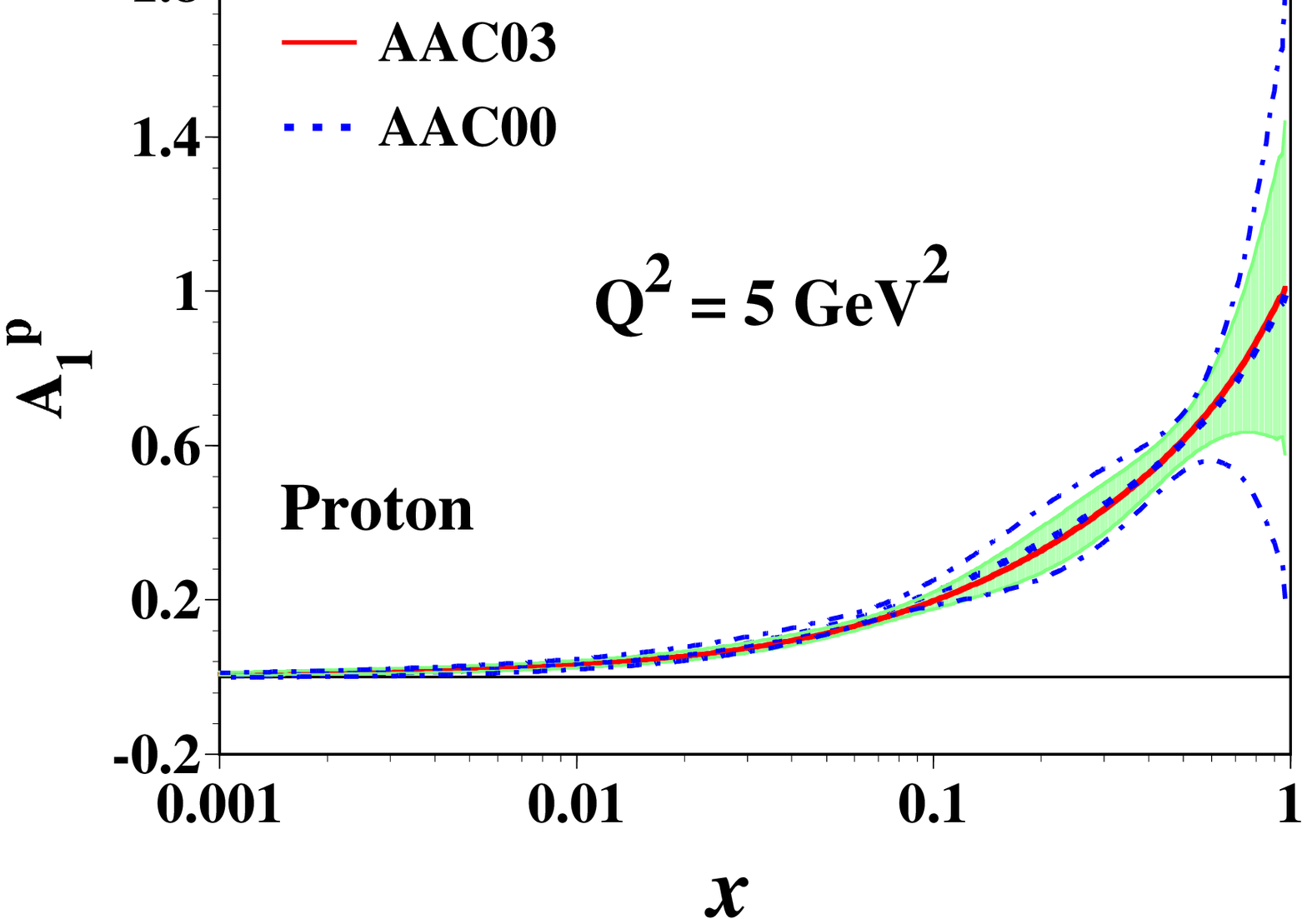} \\ \vspace{1mm}
      \includegraphics*[width=60mm]{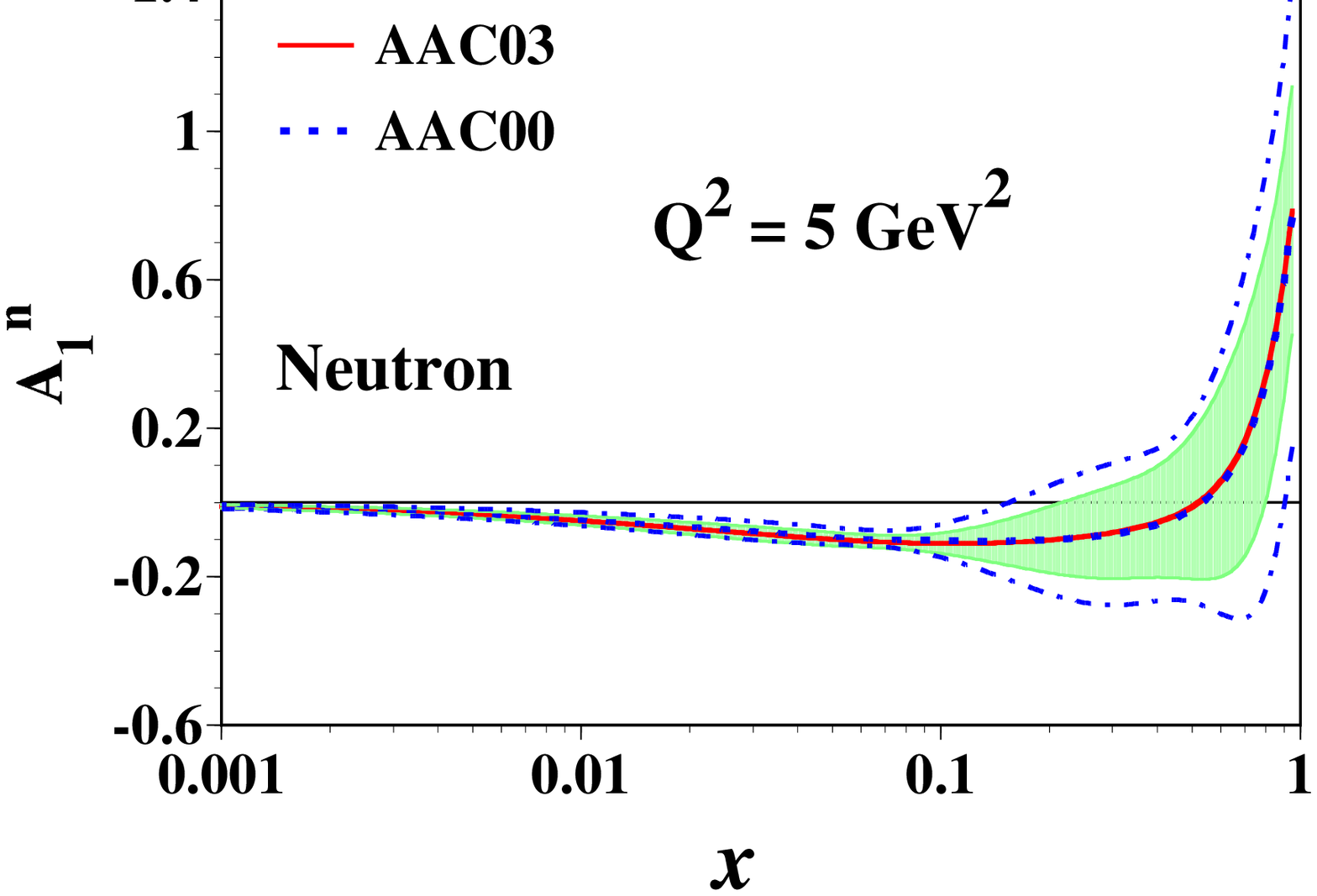} \\ \vspace{1mm}
      \includegraphics*[width=60mm]{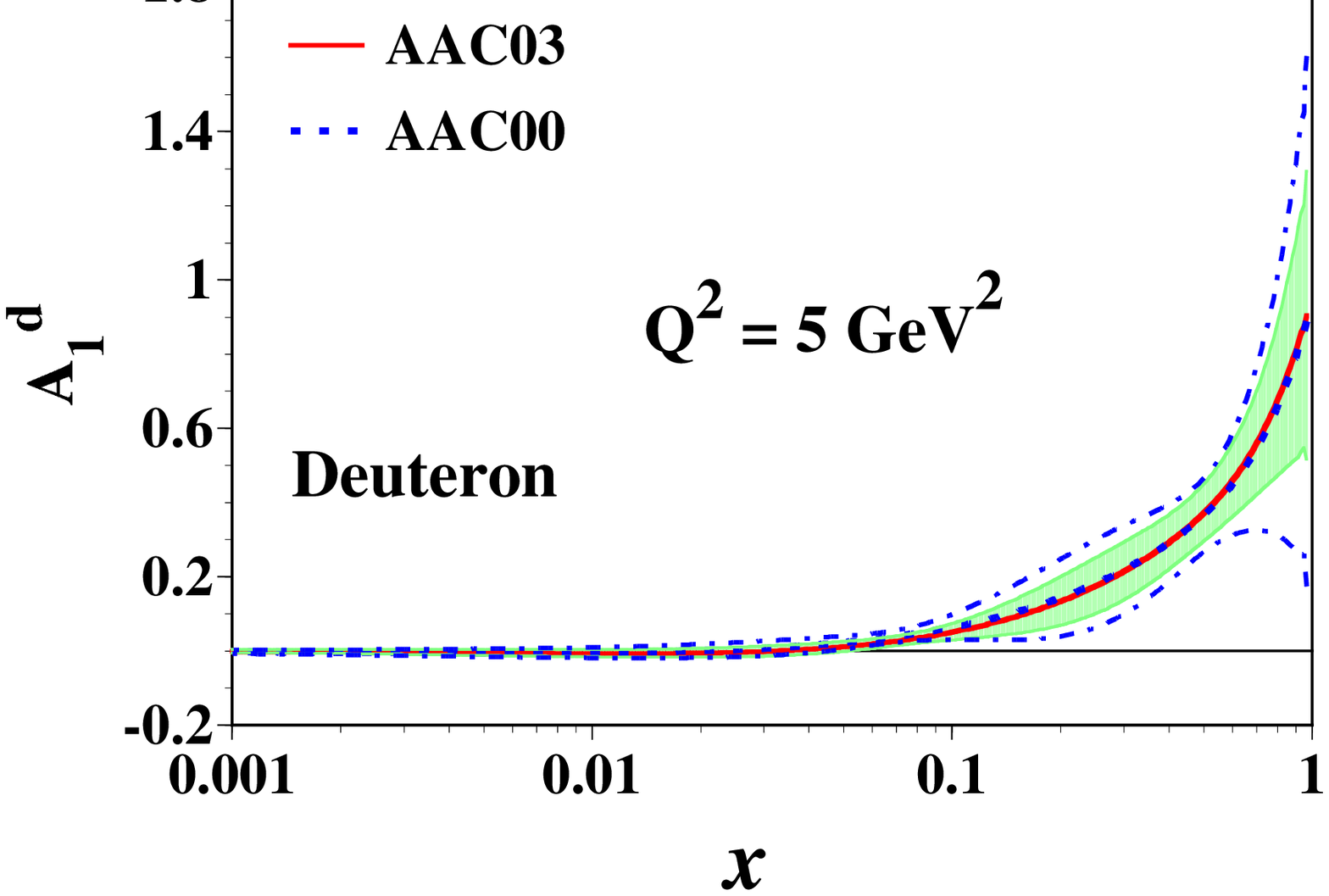} 
\caption{\label{fig:A1uncer}
Calculated spin asymmetries and their uncertainties are shown
at $Q^2=5$ GeV$^2$. The solid curves and shaded areas indicate
the spin asymmetries and their uncertainties of the new results (AAC03),
respectively. The dashed and dashed-dot curves show those of the previous
results (AAC00 NLO-2).
}
\end{figure}

\begin{figure*}
        \includegraphics*[width=42mm]{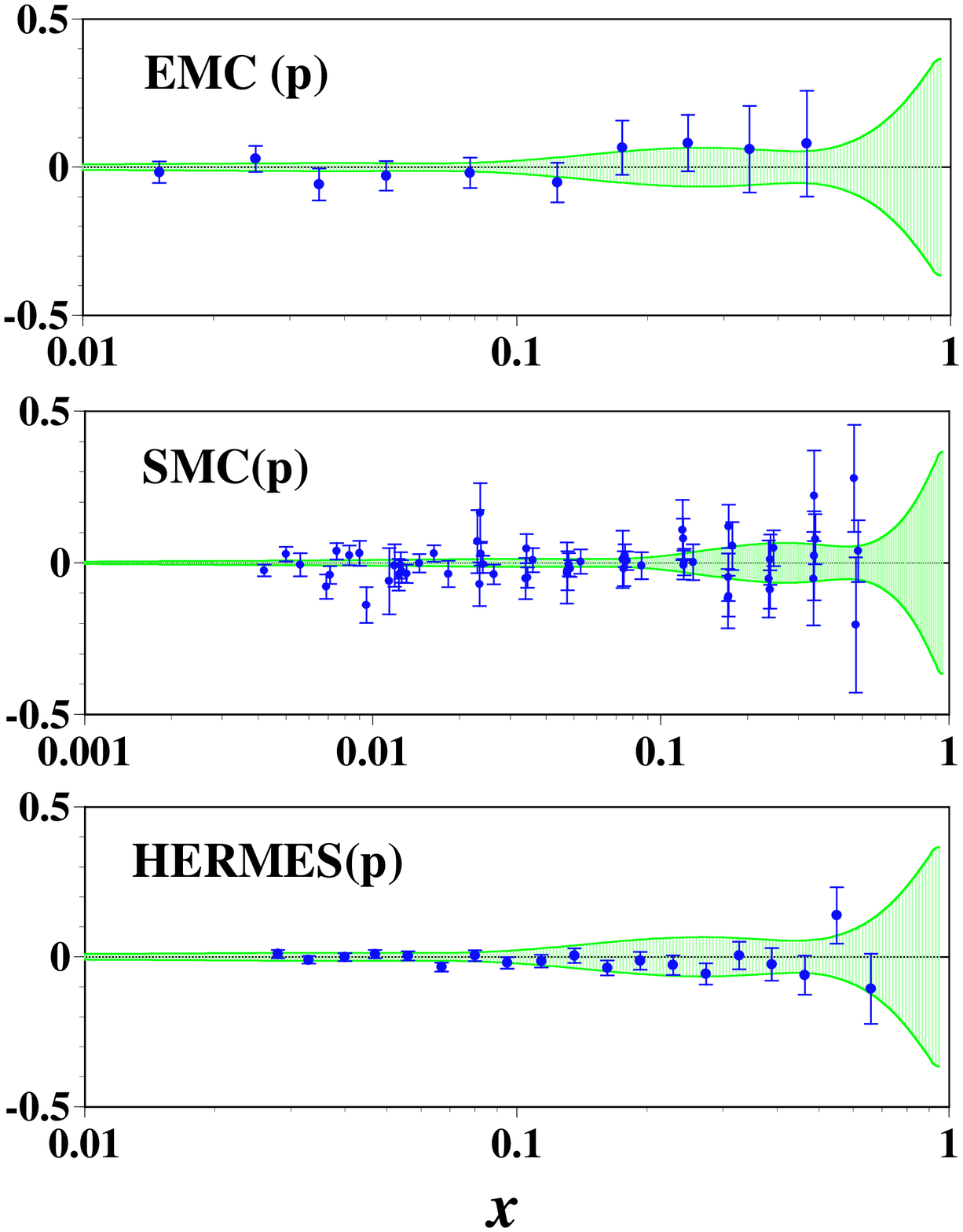} \hspace{0.5mm}
        \includegraphics*[width=42mm]{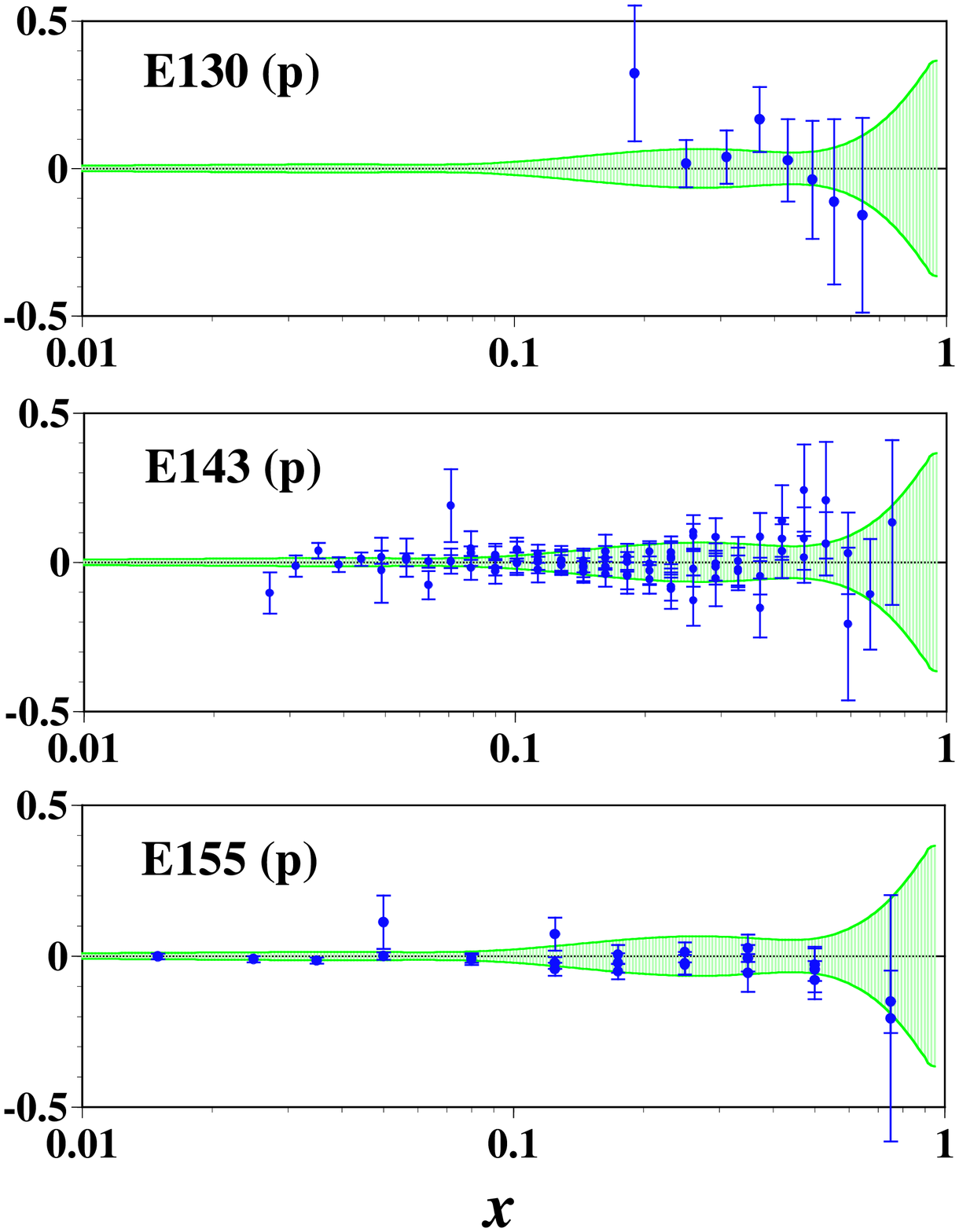} \hspace{0.5mm} 
        \includegraphics*[width=42mm]{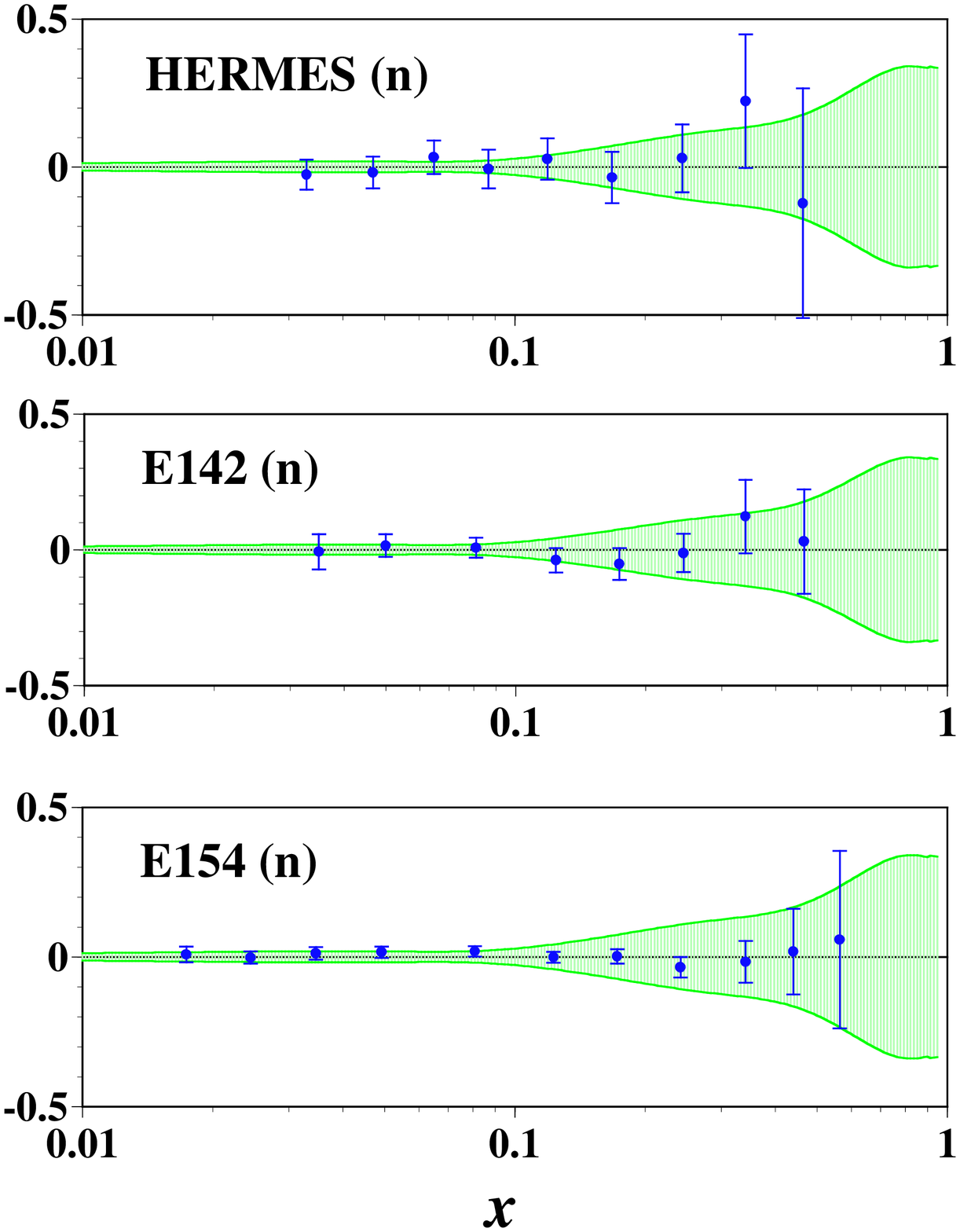}  \hspace{0.5mm}
        \includegraphics*[width=42mm]{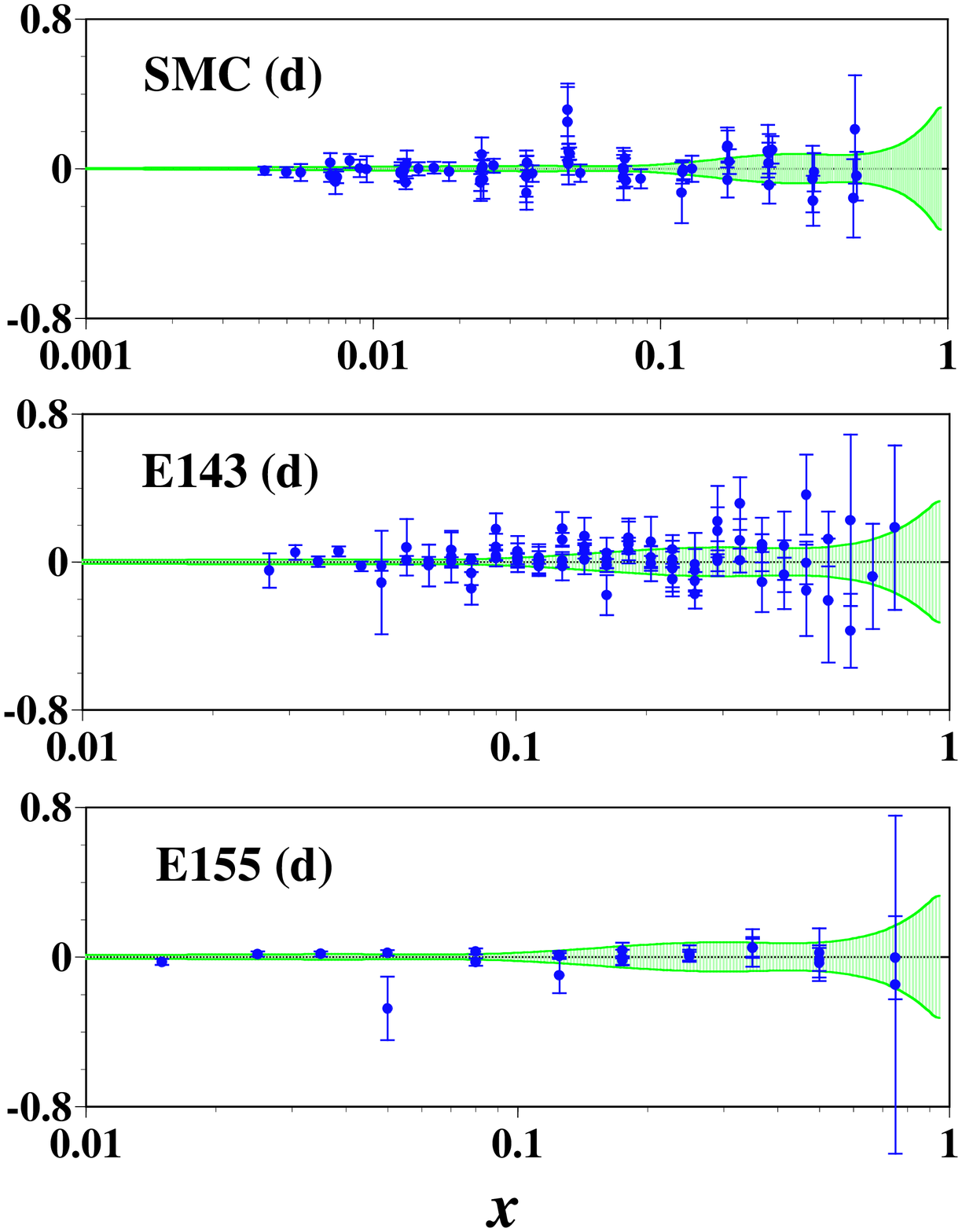} 
\caption{\label{fig:A1pexp}
Comparison of the AAC03 spin asymmetries with experimental data. 
The ordinates indicate the differences between experimental data and
theoretical values ($A_1^{data}-A_1^{AAC03}$). The error bars indicate
the errors obtained by the quadratic summations of the statistical
and systematic errors. The shaded areas show the uncertainties
at $Q^2=5$ GeV$^2$.
}
\end{figure*}

We discuss the results for the spin asymmetries and their uncertainties. 
In addition to the data used for the AAC00 analysis,
the SLAC-E155 proton target data are included. The E155 data 
cover the region, $0.015<x<0.75$ and $1.22<Q^2<34.72$ GeV$^2$.
Calculated spin asymmetries are shown in Fig. \ref{fig:A1uncer},
and they are compared with the previous results (AAC00 NLO-2).
The solid curves and shaded areas show the spin asymmetries $A_1^p$,
$A_1^n$, and $A_1^d$ and their uncertainties of the new results (AAC03),
respectively. The dashed and dashed-dot curves indicate those of
the AAC00 NLO-2. 
The $A_1$ uncertainties are calculated by using the estimated
uncertainties for the obtained polarized PDFs. 

The spin asymmetries are slightly modified especially
in the region $0.02<x<0.3$. It is noteworthy that the E155 proton
data also affect the spin asymmetries of the neutron and deuteron since
$\Delta u_v(x)$ and $\Delta \bar{q}(x)$ are modified. 
Although the asymmetry modifications are rather small, 
the uncertainties are significantly modified. 
Comparing the shaded areas with the dashed-dot curves, 
we find that the addition of the E155 data
reduces the $A_1$ uncertainties. 
The uncertainties in the region $x<0.6$ are reduced directly by
the E155 data. In addition, the data indirectly contribute to 
the uncertainty reduction in the large-$x$ region, where precise data
are not available, through the $x$-dependent PDF form.

Next, the differences between the $A_1$ data and the theoretical
asymmetries, namely $A_1^{data}-A_1^{AAC03}$, are shown 
in Fig. \ref{fig:A1pexp}. The shaded areas indicate the uncertainties
of the AAC03 analysis. The error bars indicate the ones obtained by
the quadratic summations of the statistical and systematic errors.
We find that the uncertainties are roughly equal to the errors
of the experimental data.
The spin asymmetries are constrained in the region $x<0.1$; 
however, they still have rather large uncertainties in the larger-$x$
region. It is obvious that the large-$x$ asymmetries are not 
determined well from the present data.
These uncertainties at large $x$ imply that the positivity of the spin
asymmetry is not necessarily guaranteed in the $\chi^2$ analysis
unless the positivity condition is enforced.
The shaded areas spread out in the region $0.1 <x < 0.6$
due to large errors of the E143 and SMC data for the proton.
The numbers of these data are larger than those of HERMES and E155
experiments, so that their overall $\chi^2$ contributions are larger
and the accurate E155 data cannot contribute much in this region.
The situation is similar for the deuteron uncertainties.
The neutron uncertainties are still large because the used
$^3$He target data are not accurate enough in comparison with
the proton and deuteron data. We expect that the neutron uncertainties
could be improved by precise JLab measurements \cite{JLab-n}.

\subsection{\label{PDF} 
Polarized parton distribution functions}

\begin{figure}[b]
        \includegraphics*[width=40mm]{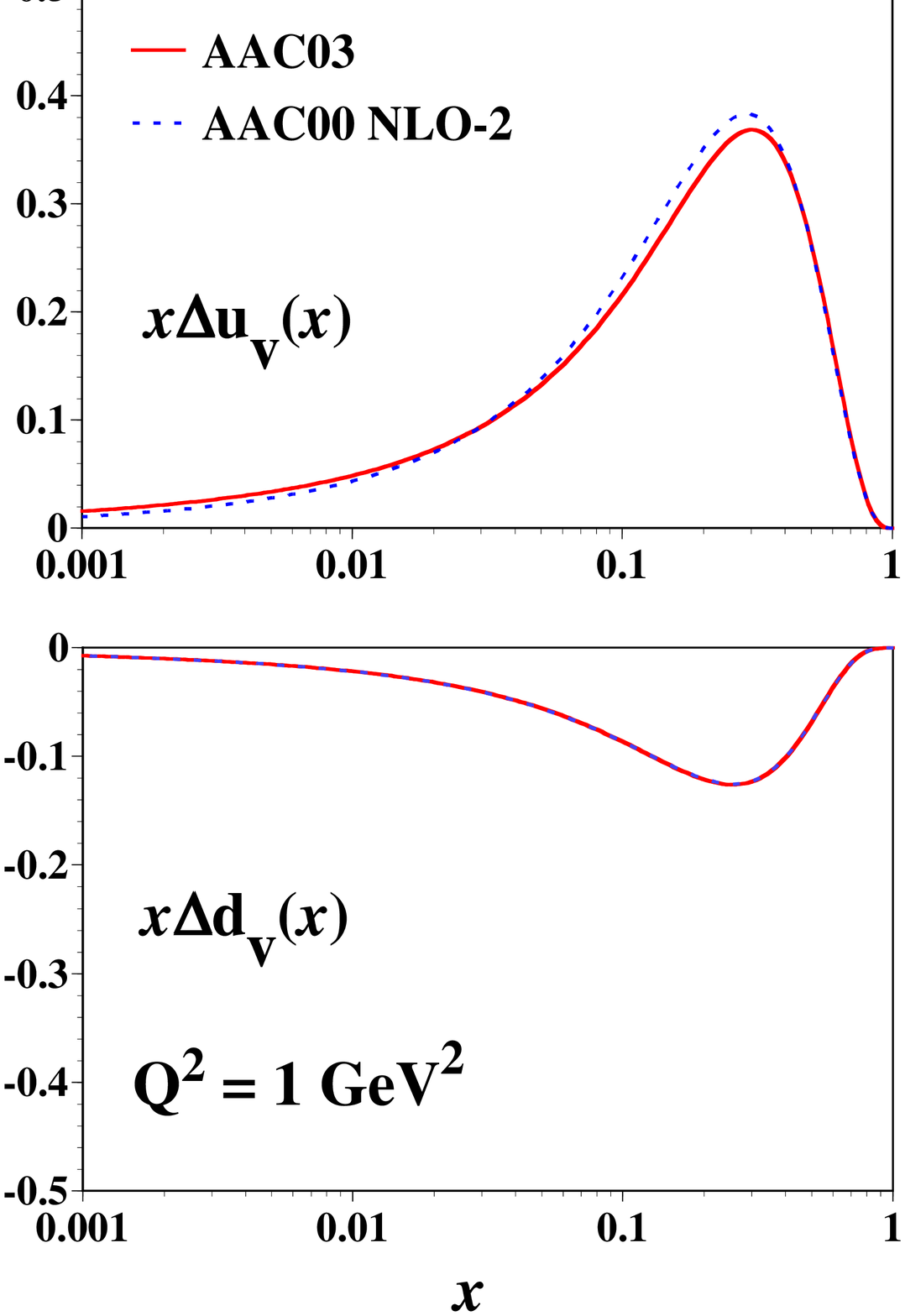} \hspace{1mm}
        \includegraphics*[width=41mm]{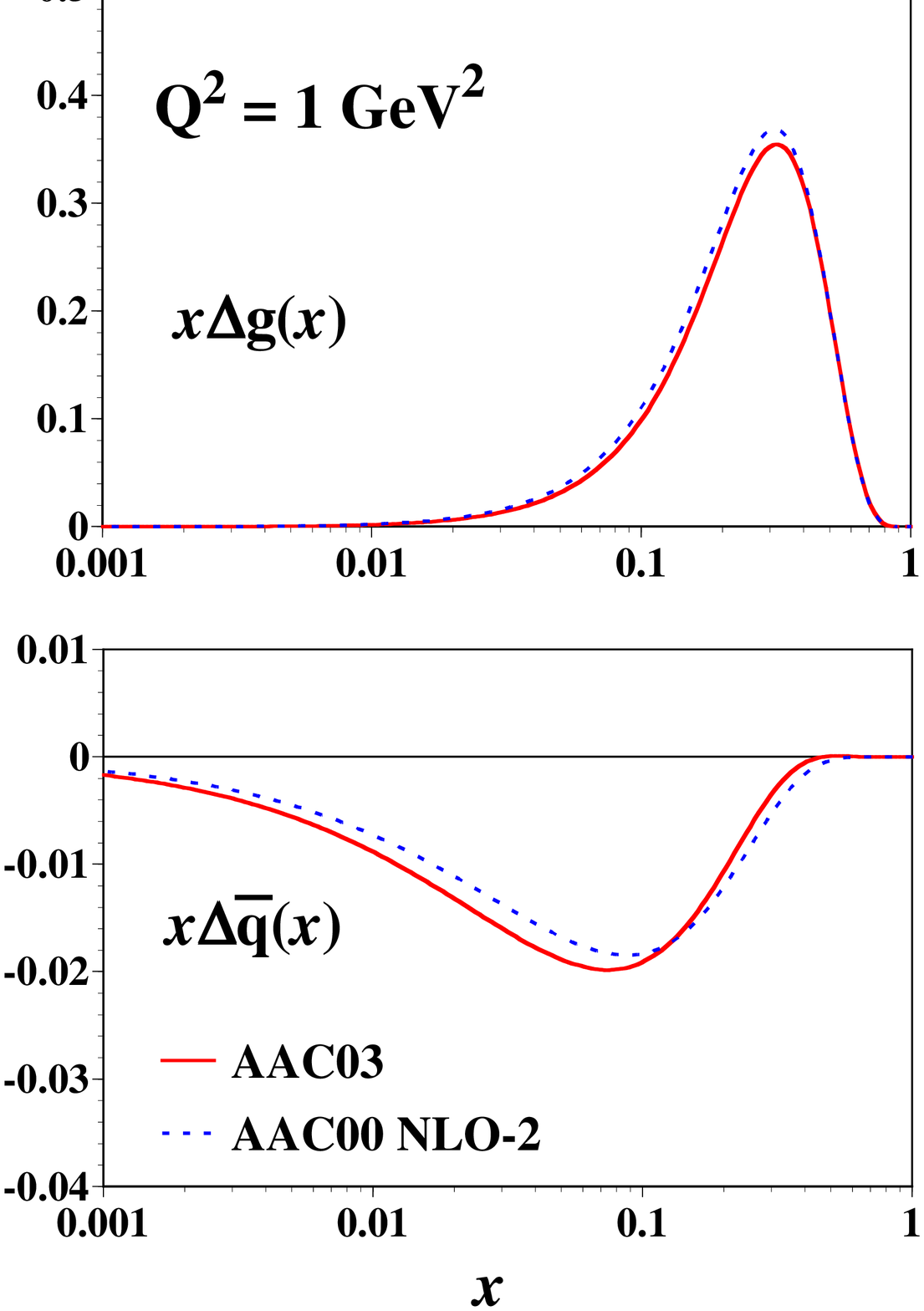}
\caption{\label{fig:xdf}
Obtained polarized parton distributions at $Q^2=1$ GeV$^2$. 
The solid curves indicate the new AAC03 results, and the
dashed curves are taken from the previous analysis (AAC00 NLO-2).}
\end{figure}

\begin{figure}[t]
        \includegraphics*[width=40mm]{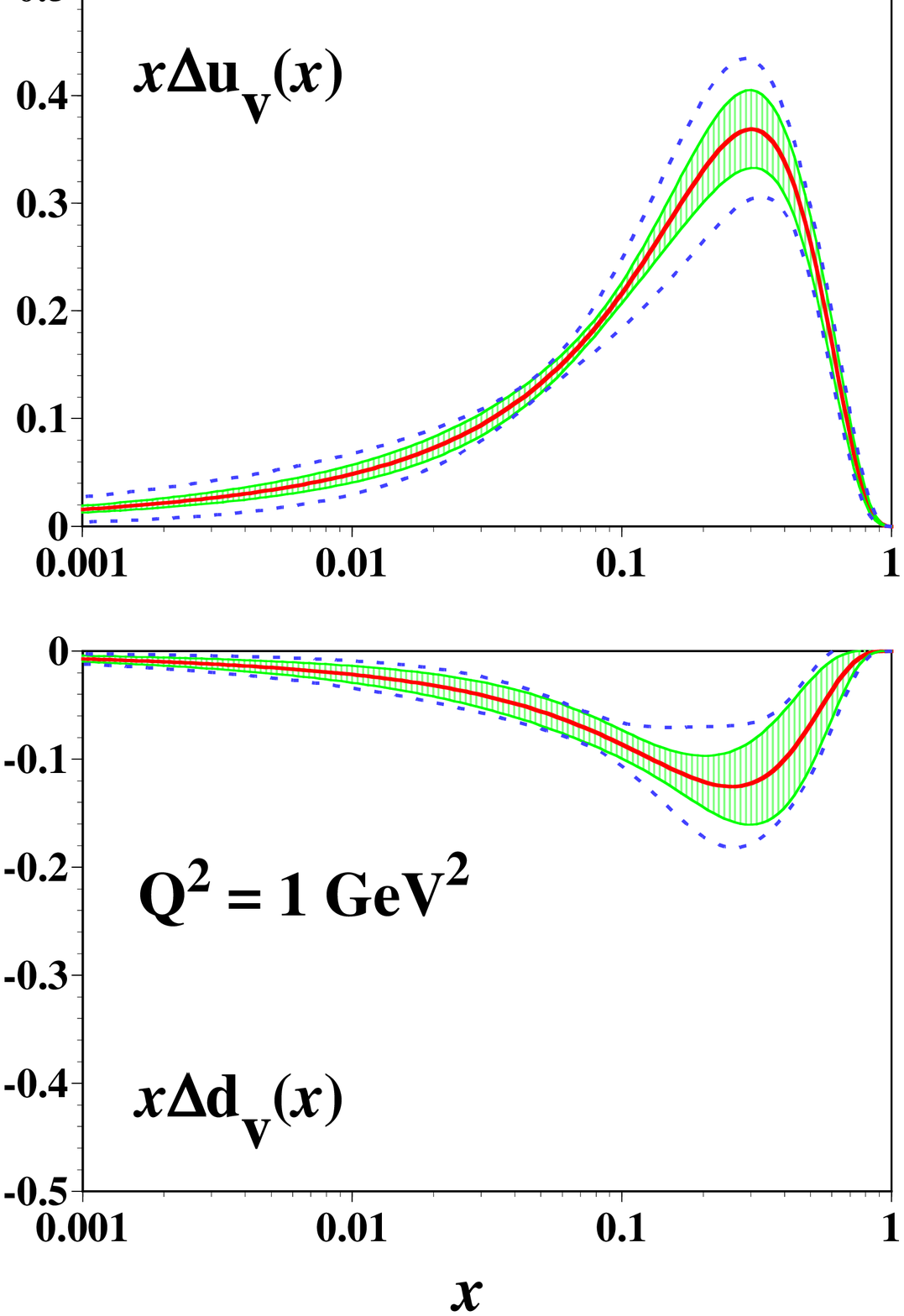} \hspace{1mm}
        \includegraphics*[width=41mm]{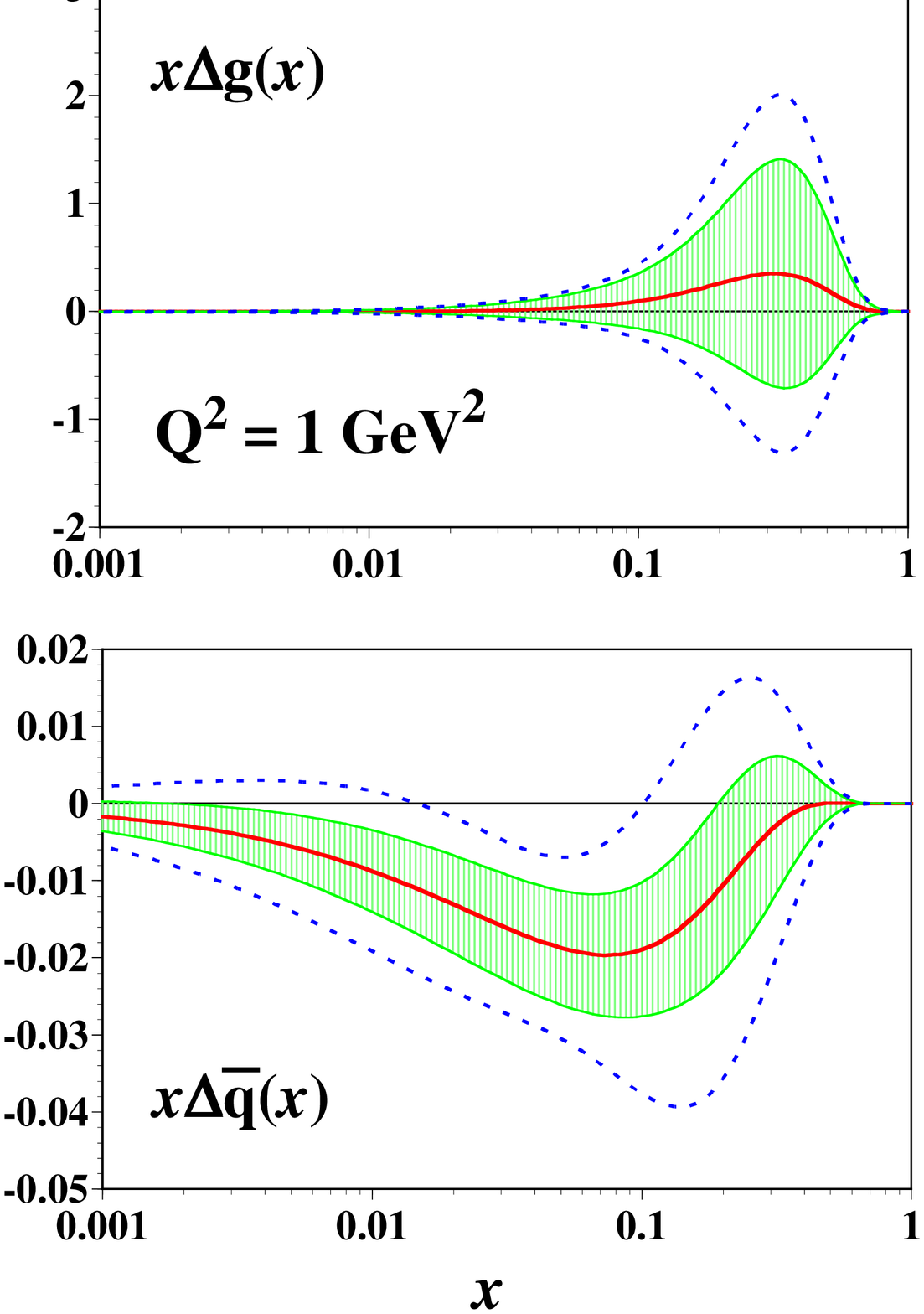}
\caption{\label{fig:exdf}
Polarized PDF uncertainties are shown at $Q^2=1$ GeV$^2$.
The solid curves and shaded areas are the polarized PDFs 
and their uncertainties of the new AAC03 results, and
the dashed curves are the uncertainties of the AAC00 results.
}
\end{figure}

We discuss obtained polarized PDFs and their uncertainties.
In Fig. \ref{fig:xdf}, the polarized PDFs of the AAC03 analysis are
compared with those of the with the AAC00 NLO-2 at $Q^2=$ 1 GeV$^2$.
The distributions $\Delta u_v$, $\Delta \bar{q}$, and $\Delta g$
are slightly modified, but $\Delta d_v$ is scarcely changed.
This is because $\Delta u_v$ is the dominant component of the proton
structure function $g_1^p$ and because the whole sea-quark contribution
is in general larger than the $\Delta d_v$ contribution.
Therefore, the larger components $\Delta u_v$ and $\Delta \bar{q}$
are mainly affected by the added precise E155 data.

Next, PDF uncertainties are shown at $Q^2=$ 1 GeV$^2$ in
Fig. \ref{fig:exdf}. The shaded areas are the uncertainties
of the AAC03 analysis, and the dashed curves indicate those for
the AAC00. The valence-quark distributions are sufficiently
constrained by the polarized DIS data. However, we find rather large
uncertainties in the region $0.1<x<0.6$, which corresponds to the
region of the large $A_1$ uncertainties in Fig. \ref{fig:A1pexp}.
It indicates the necessity of accurate data in this kinematical
region for better determination of the polarized valence-quark
distributions.
In particular, accurate $^3$He data are useful for reducing the $\Delta d_v$
uncertainties because the $\Delta d_v$ contribution to $g_1^n$
is almost the same as the $\Delta u_v$
($g_1^n \propto 4 \Delta d_v+\Delta u_v+12\Delta \bar{q}$),
whereas the contribution is small in $g_1^p$.
On the other hand, the uncertainties of the antiquark and gluon
distributions are still large. The huge gluon uncertainties 
indicate that the present data cannot rule out the possibility of
$\Delta g(x)=0$ and negative gluon polarization,
although the obtained gluon distribution is positive.

As shown in Fig. \ref{fig:exdf}, all the PDF uncertainties are
significantly reduced in the AAC03 analysis in comparison with the AAC00
because the accurate E155 data are added to the data set.
In particular, the $\Delta d_v$ uncertainties are reduced
although the $\Delta d_v$ distribution stays almost the same.
In addition, the uncertainties of the antiquark and gluon
distributions are significantly improved. The antiquark uncertainty
reduction is directly due to the E155 data.
However, it is difficult to understand that the significant reduction
of the gluon uncertainties is due to the added new data.
This is because that the gluon distribution indirectly contributes
as a higher order correction with the coefficient function,
and this contribution is less than quark contributions. 
The huge gluon uncertainties explicitly indicate
the difficulty of fixing the gluon distribution from the DIS
experimental data. 

We find that the gluon uncertainty reduction is caused by an error
correlation.
The non-diagonal part of the Hessian indicates a strong correlation
between the polarized antiquark and gluon distributions.
The correlation affects the determination of these distributions.
We discuss the details of the uncertainty improvement
due to the correlation in Sec. \ref{correlation}.

\subsection{\label{1stm} Quark spin content}

We show the first moments of the AAC03 parameterization
at $Q^2=$ 1 GeV$^2$ in Table \ref{T:1stm}, and they are compared with
those of the AAC00 NLO-2 set. The first moments of the up-
and down-valence quark distributions are fixed in both analyses. 
The first moments indicate that quarks carry about 20$\%$ of the parent
nucleon spin, and gluons carry a large positive fraction of the nucleon
spin. Their uncertainties are significantly reduced by the added E155 data; 
however, the present data are not enough to obtain accurate values,
especially for the gluon first moment. 

The uncertainty of the spin content $\Delta \Sigma$ strongly depends
on the antiquark uncertainty because it is given
by $\Delta \Sigma_{N_f=3}=\Delta u_v+\Delta d_v+6\Delta \bar{q}$. 
The first moments of the valence-quark distributions are fixed,  
so that the $\Delta \Sigma$ uncertainty is equal to six times
the $\Delta \bar{q}$ uncertainty, which could be large due to
the uncertainty of the distribution $\Delta \bar{q}(x)$
in the small-$x$ region.
It suggests that the extrapolation into the smaller-$x$ region
should be ambiguous in calculating the integral over $x$.
We expect that accurate polarized antiquark distributions will be measured
in future, then the quark spin content issue will become clear.

\begin{table}[t]
\caption{\label{T:1stm}
The first moments of the obtained polarized PDFs at $Q^2=1$ GeV$^2$. 
The AAC03 analysis results are compared to those of the previous
results (AAC00 NLO-2). The $\Delta \Sigma$ is the quark spin content.
}
\begin{ruledtabular}
\begin{tabular}{lccc} 
        & $\Delta \bar{q}$     & $\Delta g$        &$\Delta \Sigma$     \\
\hline
AAC03   & $-$0.062 $\pm$ 0.023 & 0.499 $\pm$ 1.266 & 0.213 $\pm$ 0.138  \\
AAC00   & $-$0.057 $\pm$ 0.037 & 0.533 $\pm$ 1.931 & 0.241 $\pm$ 0.225 
\end{tabular}
\end{ruledtabular}
\end{table}

\subsection{\label{otherPDFs} Comparison with other parameterizations}

\begin{figure}[b]
        \includegraphics*[width=40mm]{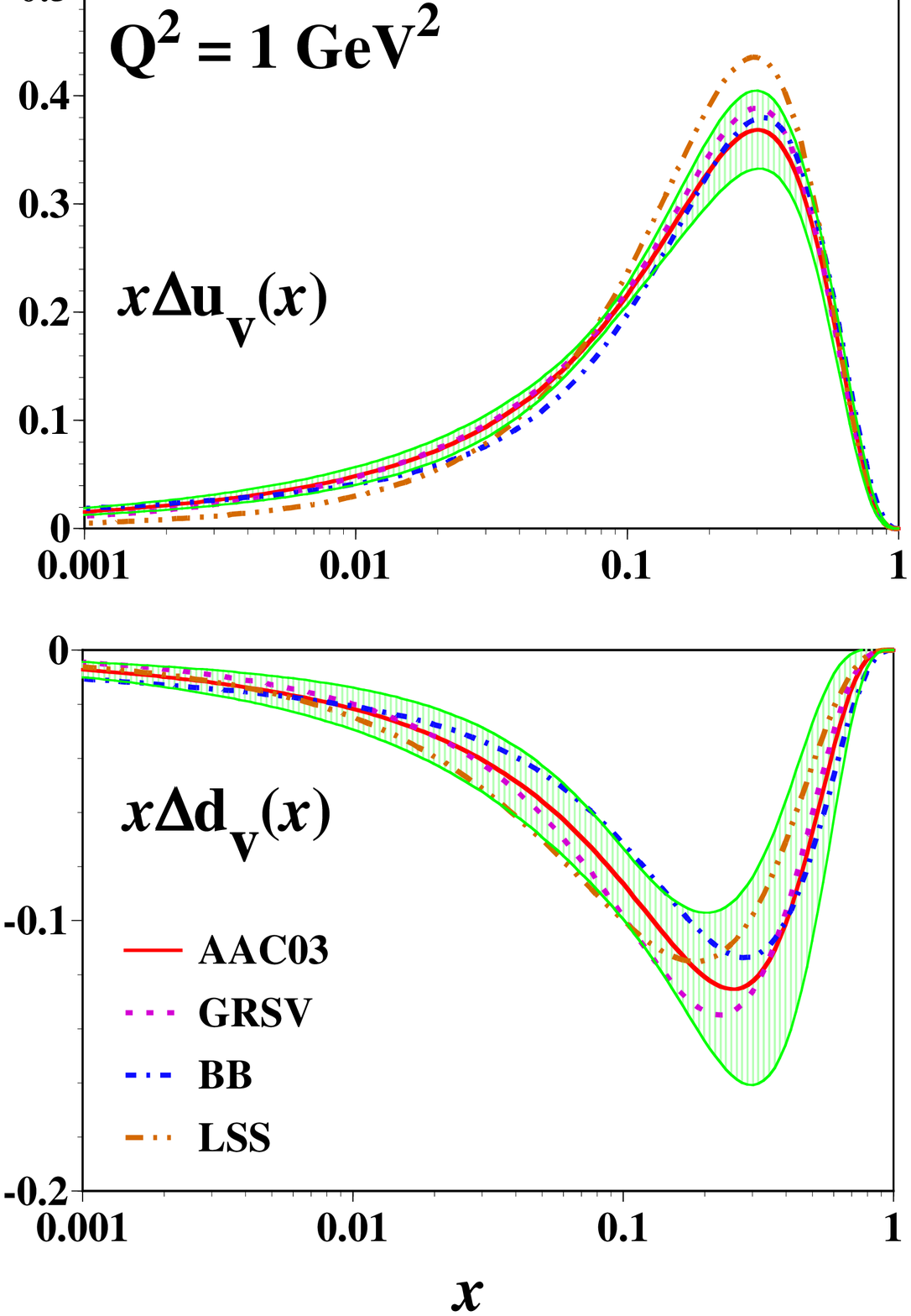} \hspace{1mm}
        \includegraphics*[width=40mm]{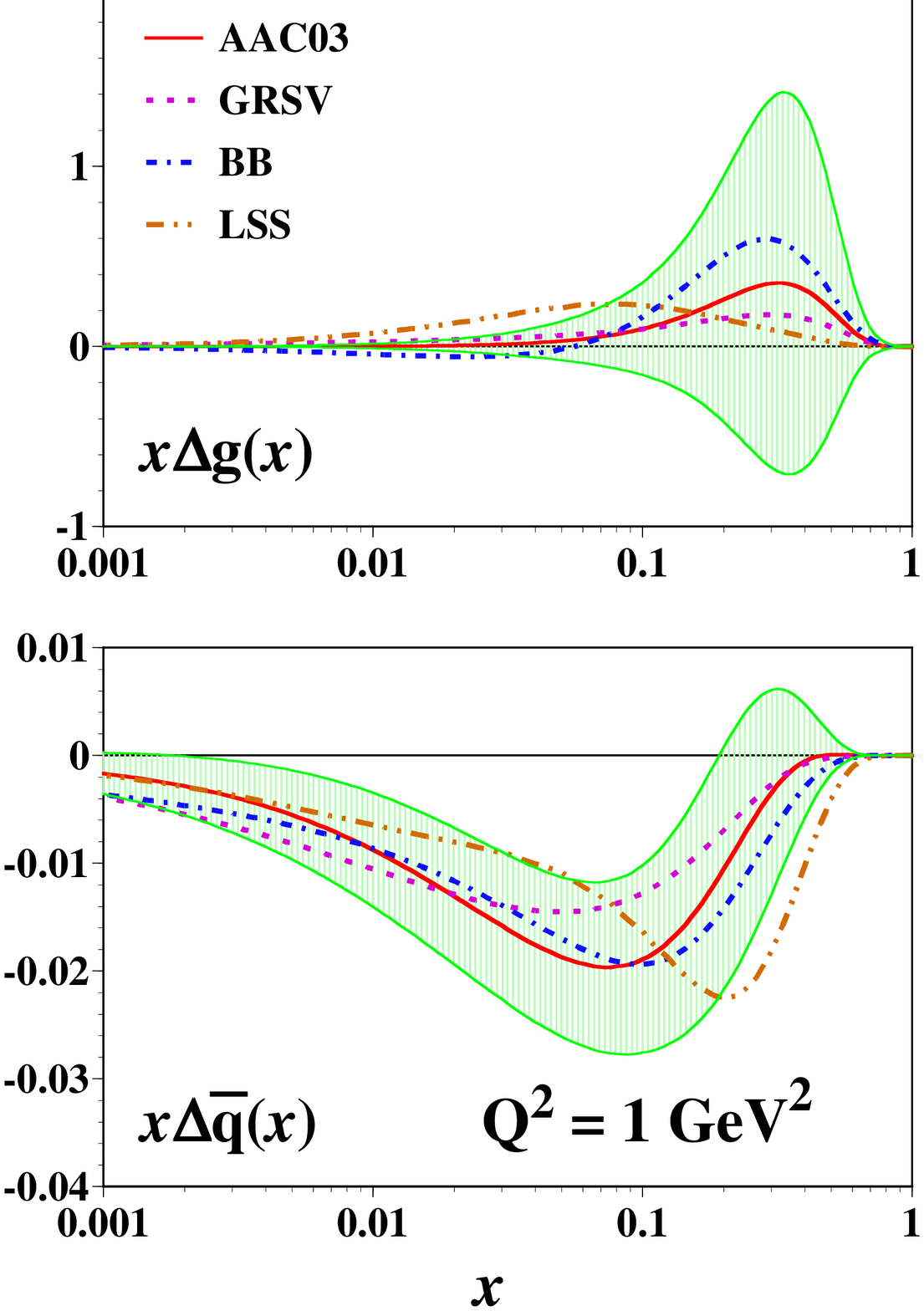}
\caption{\label{fig:vsother}
The AAC03 PDFs at $Q^2=1$ GeV$^2$ are compared with the ones for
other parameterizations by GRSV01 (standard scenario) \cite{GRSV},
BB (ISET=3) \cite{BB}, and LSS ($\overline{\rm MS}$ scheme) \cite{LSS}.
The shaded areas are the uncertainties of the AAC03 analysis.
}
\end{figure}

The AAC03 analysis results are compared with other parameterizations
at $Q^2=1$ GeV$^2$ in Fig. \ref{fig:vsother}.
For comparison, we choose three sets of the polarized PDFs in the NLO: 
GRSV01 (standard scenario) \cite{GRSV}, BB (ISET=3) \cite{BB}, 
and LSS ($\overline{\rm MS}$ scheme) \cite{LSS}.
These parameterizations used basically the same experimental data set
of the polarized DIS, but they choose averaged data tables
over $x$ and $Q^2$, whereas full tables are used in the AAC analysis.
The flavor symmetric antiquark distributions are used in all these
parameterizations.

Because the first moments are fixed in the same way, the variations
are small in the polarized valence-quark distributions
among the parameterizations in Fig. \ref{fig:vsother}.
There are slight variations in the antiquark distributions, and
the gluon distributions differ significantly among
the analysis groups. However, we find that all the parameterizations
are consistent each other because the distributions are mostly within
the estimated error bands. 

The BB and LSS groups also investigated the polarized PDF uncertainties 
by the Hessian method. However, these uncertainties may not be directly
compared with our uncertainties because the used $\Delta \chi^2$ value
is not specified in their papers. In addition, uncertainty estimation
methods are slightly different; for example, only the statistical errors
are used and a relative normalization shift is introduced in the BB analysis.
In general, the error estimations involve complicated systematic errors,
{\it e.g.} functional form, data selection, and higher-twist effects,
in the global analyses, and they may not be estimated numerically.
It is difficult to clarify these issues only by the current polarized
DIS experiment data. Therefore, we need to investigate such hidden
uncertainties when we improve the quality of the polarized PDFs 
by incorporating future experimental data.

\subsection{\label{correlation} 
$\Delta g (x)=0$ analysis and error correlation}

\begin{figure}[b]
        \includegraphics*[width=40mm]{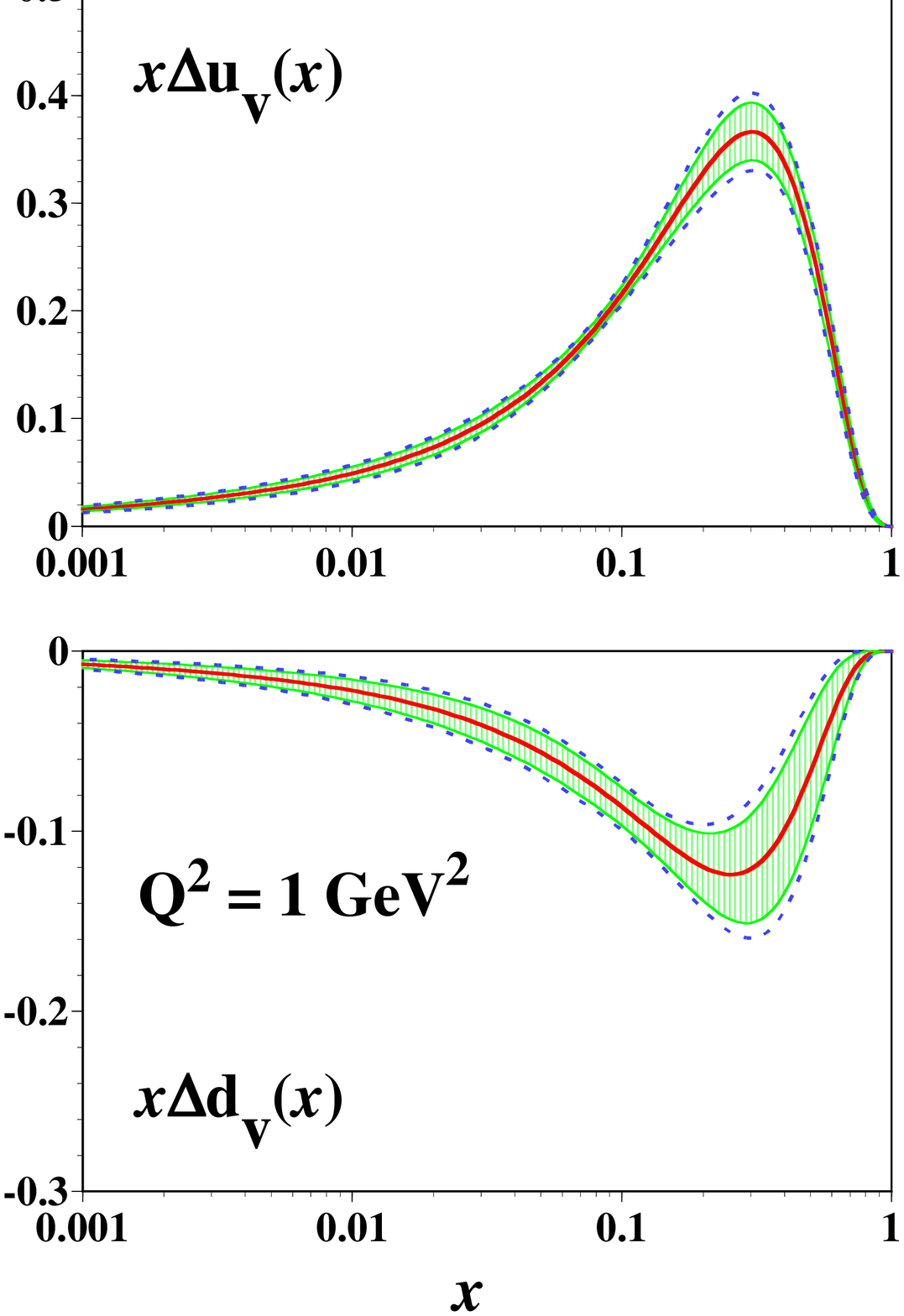} \hspace{1mm}
        \includegraphics*[width=41mm]{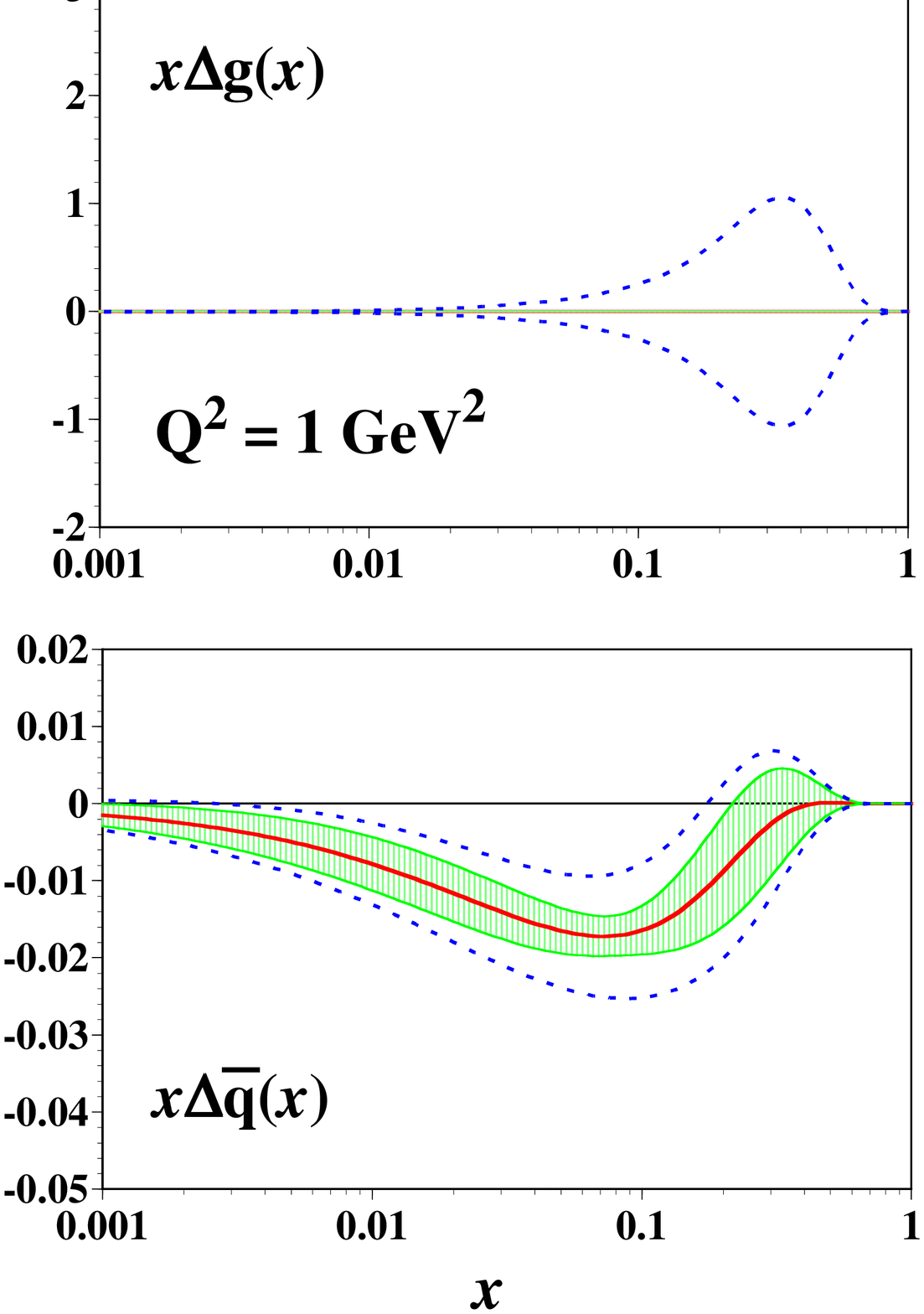}
\caption{\label{fig:exdf-g0}
The PDF uncertainties of the $\Delta g(x) = 0$ analysis are compared 
with those of the $\Delta g(x) \neq 0$ analysis (AAC03).
The solid curves indicate the polarized PDFs of the $\Delta g(x)=0$ analysis
at $Q^2=1$ GeV$^2$, and the shade areas are their uncertainties.
The dashed curves indicate the uncertainties of the $\Delta g(x) \neq 0$
analysis.
}
\end{figure}

In order to understand the reduction of the gluon uncertainty
in Fig. \ref{fig:exdf}, we discuss the error correlation between
the polarized antiquark and gluon distributions. As the most simple
assumption which could be allowed within the gluon uncertainties
in Fig. \ref{fig:exdf}, we choose $\Delta g(x)=0$ at the initial
scale ($Q^2=1$ GeV$^2$). However, one should note that a finite
distribution $\Delta g(x) \ne 0$ appears at larger $Q^2$ from
the singlet $Q^2$ evolution. Since the gluon-distribution parameters
are fixed, we can carry out an uncorrelated analysis with
the gluon distribution.

In the $\Delta g(x)=0$ analysis, we obtain 
$\chi^2 \ (/d.o.f.)$$=355.0$ (0.915),
which is $2.5 \%$ larger than the value for the $\Delta g(x) \neq 0$ analysis. 
Because it is a slight change in the $\chi^2$ value, it is reasonable that
$\Delta g(x) =0$ is allowed in Fig. \ref{fig:exdf} if
the uncertainties are taken into account. Obtained polarized PDFs are shown
in Fig. \ref{fig:exdf-g0} for the $\Delta g(x)=0$ analysis.
The total number of optimized parameters is seven for this analysis,
so that the uncertainties are estimated by $\Delta \chi^2=8.180$.
The calculated uncertainties are shown by the shaded areas, and they
are compared with those of the $\Delta g(x) \neq 0$ analysis (AAC03)
shown by the dashed curves. We find that the antiquark uncertainties
are significantly reduced. On the other hand,
the valence quark uncertainties are scarcely changed, which indicates
that the correlation with the gluon distribution is weak even in
the $\Delta g(x) \ne 0$ analysis.

It suggests that the antiquark distributions should be determined well
by the present data if their errors are uncorrelated with those of
the gluon. However, because of the existence of the strong 
error correlation, the actual antiquark uncertainties are increased
by the huge gluon uncertainties.
In future, if the polarized gluon distribution is measured accurately,
the antiquark uncertainties also become small due to
the strong correlation. To the contrary, the gluon uncertainties
could be reduced by accurate measurements of the antiquark
distributions. In this way, we find the significant reduction of
the antiquark uncertainties in Fig. \ref{fig:exdf-g0} is caused
by the error correlation effects with the polarized gluon distribution.
Furthermore, it indicates that the gluon uncertainty reduction
in Fig. \ref{fig:exdf} is also due to the correlation effect between
these distributions.

From these studies, it becomes clear that accurate determination
of the gluon distribution is important also for the determination of
the antiquark distributions. At this stage, the polarized gluon
distribution is not accurately determined, and it also makes it difficult
to fix the antiquark distributions from the DIS experimental data.
In this sense, it is important to measure the polarized gluon
distribution, for example, by direct photon production
and jet production at RHIC.

\section{\label{Summary} Summary}

We have investigated the optimum polarized parton distributions by analyzing
the polarized DIS data. We focused our studies particularly on three aspects,
the uncertainty estimation of the obtained PDFs, the role of the accurate
E155 proton data, and the error correlation between the polarized gluon
and quark distributions. 

First, the obtained PDF uncertainties indicated that
the polarized valence-quark distributions are determined well,
that the uncertainties  of the polarized antiquark distributions
are slightly large, and that the gluon uncertainties are huge.
It is obvious that the polarized gluon distribution cannot be
determined from the present DIS data.

Second, we discussed the role of accurate E155 data in the global
analysis. Comparing the AAC00 and AAC03 results, we clarified that
the E155 data contributed to reducing the PDF uncertainties significantly.

Third, the error correlation is investigated by repeating
the parametrization analysis with the initial condition
$\Delta g(x)=0$. In the $\Delta g(x)=0$ analysis, there is no error
correlation between the gluon and quark distributions, and it leads
to small uncertainties of the obtained antiquark distributions.
This fact suggests that precise gluon measurements should be
valuable for a better determination of the polarized antiquark
distributions. The opposite way is also right: Precise polarized
quark measurements should provide constraints for the gluon 
distribution. 

Finally, we mention that the AAC03 polarized PDF library is available
at the web site \cite{aacweb}. The polarized PDFs can be calculated
numerically at given $x$ and $Q^2$ values.

\begin{acknowledgements}

The authors thank the AAC members, especially Y. Goto, H. Kobayashi,
M. Miyama, and T.-A. Shibata, for discussions.
S.K. was supported by the Grant-in-Aid for Scientific Research
from the Japanese Ministry of Education, Culture, Sports, Science,
and Technology. This work is partially supported by the
Japan-U.S. Cooperative Science Program. 
\end{acknowledgements}


\end{document}